\documentclass[prd,showpacs,amsmath,amssymb,preprint,nofootinbib]{revtex4-1}
\def\br{\begin{eqnarray}}
\def\er{\end{eqnarray}}
\def\be{\begin{equation}}
\def\ee{\end{equation}}

\def\l{\lambda}

\def\({\left(}
\def\){\right)}
\def\s{\sigma}

\def\lesssim{\mathrel{\hbox{\rlap{\hbox{\lower4pt\hbox{$\sim$}}}\hbox{$<$}}}}
\def\gtrsim{\mathrel{\hbox{\rlap{\hbox{\lower4pt\hbox{$\sim$}}}\hbox{$>$}}}}
\makeindex
\begin{document}

\title{A minimal 3-3-1 model with naturally sub-eV neutrinos}

\author{C. A. de S. Pires, F. S. Queiroz, P. S. Rodrigues da Silva}
%\email{farinaldo@fisica.ufpb.br,cpires@fisica.ufpb.br, psilva@fisica.ufpb.br}
\address{{ Departamento de
F\'{\i}sica, Universidade Federal da Para\'\i ba, Caixa Postal 5008, 58051-970,
Jo\~ao Pessoa, PB, Brasil}}

%\date{\today}

\begin{abstract}
In the original version of the minimal $SU(3)_C \times SU(3)_L \times U(1)_N$ model the masses of all quarks are correctly obtained by introducing three scalar triplets into the model, meanwhile the lepton mass generation requires the introduction of at least one scalar sextet. In this work we show that this scalar sextet is unable to yield the correct neutrino masses and mixing. In order to solve this puzzle in the most economical way, we evoke an additional $Z_3$ discrete symmetry, without including this sextet in the scalar spectrum, and propose a truly minimal 3-3-1 model capable of generating the correct masses and mixing of all fermions. Moreover, we show that our proposal leads to naturally light neutrinos with masses in the eV range, obtained with three scalar triplets only. Finally, the so called minimal 3-3-1 model is also in danger due to the presence of undesirable effective operators that lead to proton decay unless they are suppressed by extremely small couplings or, as we choose to employ in this work, are eliminated by some discrete symmetry.
\end{abstract}

%\pacs{ 14.60.St; 14.60.Pq; 12.60.Cn; 12.60.Fr.}
%

%
%\PACS 14.60.St; 14.60.Pq; 12.60.Cn; 12.60.Fr.

\maketitle
\section{introduction}
Although the standard model of the electroweak interactions (SM) is remarkably successful in describing precision experiments in Particle Physics, we still have many reasons to look for Physics beyond the SM. For instance,  the SM is unable to  explain either  the observed neutrino oscillation phenomenon~\cite{osc} or the matter-antimatter asymmetry of the Universe. It also does not contain a candidate for the dark matter component of the Universe~\cite{DM}. From the theoretical side, new Physics is necessary to explain the hierarchy problem as well as  the problem of family replication, electric charge quantization, the strong CP problem, etc. It thus becomes evident that we have enough, experimental or theoretical, reasons  to go beyond the SM. 

Unfortunately, until now we have no available complete new theory that is able to account for all the experimental and theoretical problems faced by the SM. Each particular route beyond  the SM is capable of providing an explanation to a couple of such problems as, for instance, Supersymmetry theories  that can explain the hierarchy problem; Technicolor models were also proposed with the same aim, as well as large extra dimensions theories, while grand unification theories explain electric charge quantization, etc.
While there is no experimental or theoretical result that definitely allows us to discard some of these theories for not being physically realizable, all should be considered on the same footing as reasonable theoretical proposals.  Among these theories, as far as we know, the only one that  provides an explanation of the family replication are  those theories based on the $SU(3)_C \times SU(3)_L \times U(1)_N$ (3-3-1) gauge symmetry. In the framework of 3-3-1 theories, an explanation of the family replication arises because the gauge anomalies are absent only if there is a multiple of three families in their fermion spectrum, and when conjugated with asymptotic freedom, this leaves no room for more than the three fermion families. Moreover, these theories also provide a natural explanation of the strong CP problem~\cite{pal},  electric charge quantization~\cite{ECQ} and  possess genuine dark matter candidates~\cite{DM331}. 

Models for the electroweak interactions based on 3-3-1 symmetry were first intensively  explored in the 1970's~\cite{earlymodels}.  However, we stress that all those models involved exotic leptons. Actually, regarding this peculiar feature, it is necessary to say that there are two versions of 3-3-1 gauge models. In one version, the third component of the leptonic triplet is a simply charged particle. The other version involves a neutral lepton as the third component of the leptonic triplet. 

In the 1990s it was perceived that the exotic leptons could be replaced by the standard ones. More precisely, in 1992 a 3-3-1 gauge model was proposed where the third component of the lepton triplet was recognized as the anticharged lepton~\cite{ppf}. Then, in 1993, a second version of the 3-3-1 gauge model with a right-handed neutrino was insightfully proposed as the third component of the lepton triplet~\cite{footpp}. The first version~\cite{ppf} is nowadays called the minimal version of the 3-3-1 gauge model, minimal 3-3-1 for short,  because its leptonic sector is composed exclusively of the standard model leptons, while the second version~\cite{footpp} is called the 3-3-1 model with right-handed neutrinos.   

In this work we will focus on the minimal 3-3-1 model~\cite{ppf}. Our main goal is to show that the model, in its current version, cannot accommodate small neutrino masses concomitantly with the observed charged lepton masses without destroying the neutrino oscillation phenomenology. Of course, more scalar multiplets could be added to provide the correct observed data, or even the absence of the larger scalar multiplets may be considered if neutrino masses are supposed to be generated through effective operators. However, we can show that even in this case the model fails to supply an acceptable solution to this puzzle if no extra symmetry is imposed. Besides, we show that the minimal 3-3-1 model experience the undesirable presence of effective operators that lead to fast proton decay, which is due to the fact that the energy scale of its underlying theory has to be at some TeV~\cite{landaupole}.
This is not a peculiarity of this model, instead, many models that work only to some new low energy scale (mainly around a few TeV) may suffer the same drawback~\footnote{Remember that the SM is acceptable even at energies above Planck scale, and it is the presence of baryon violating effective operators that gives a clue about a lower bound on the underlying scale for the SM if only grand unified theories come next~\cite{baryonv}.}.
Here we seek the most economical solution to these problems, both achieved by the assumption of additional discrete symmetries. As a result, we will obtain small neutrino masses with no fine-tuning in the couplings and an improvement in the charged lepton Yukawa couplings, in the sense of their naturalness, when compared to the SM.

This work is organized in the following way: In Sec.~\ref{sec1} we give a short review of the minimal 3-3-1 model and argue that it is unable to provide a natural explanation of the smallness of the neutrino mass and is plagued by dangerous effective operators that could engender  fast proton decay. Next, in Sec.~\ref{sec2} we present our modification of the minimal model which solves such problems. We summarize our results in Sec.~\ref{sec3}.

\section{The minimal 3-3-1 model}
\label{sec1}
In this section we briefly present the minimal 3-3-1 model in its original version and discuss its  main problems. In supposing only the existence of standard leptons, the model mimics the structure of theories of grand  unification  because all  leptons (left-handed and right-handed) of each family come inside the same triplet representation of $SU(3)_L$,
%
%\begin{eqnarray}
\be
f_{l_L}^T = \left (
%\begin{array}{c}
\nu_l\,, % \\
e_l\,, % \\
e^{c}_l
%\end{array}
\right )^T_L\sim(1\,,\,3\,,\,0),
\label{leptoniccontent}
\ee
where $l=e,\,\mu ,\, \tau$ and $T$ means transposition. Thus, we can immediately conclude that the model may explain electric charge quantization {\it \`a la} grand unification theories~\cite{ECQ}. 
Of course, the real world demands quarks, but unfortunately the same does not happen with quarks. Concerning the quark sector, anomaly cancellation requires that one family of left-handed quarks comes in a triplet representation of $SU(3)_L$ with the following  content (below we also include right-handed quarks which are singlets),
\begin{eqnarray}
&&Q_{1_L}^T=
\left (
%\begin{array}{c}
u^{\prime}_1 \,, %\\
d^{\prime}_1 \,, %\\
J^{\prime}_1
%\end{array}
\right )_L^T \sim (\mbox{{\bf 3}},\mbox{{\bf 3}},\frac{2}{3}),\,\,\,\nonumber \\ u^{\prime}_{1_R}\sim(\mbox{{\bf 3}},&&\mbox{{\bf 1}},\frac{2}{3}),\,\,\,\,\,
d^{\prime}_{1_R}\sim(\mbox{{\bf 3}},\mbox{{\bf 1}},-\frac{1}{3}),\,\,\,\,\, J^{\prime}_{1_R}\sim(\mbox{{\bf 3}},\mbox{{\bf 1}},\frac{5}{3}),
\label{quarks1} 
\end{eqnarray}
with $u^{\prime}_{1}$  and  $d^{\prime}_{1}$ being the standard quarks, while 
$J^{\prime}_{1}$ is an exotic quark with electric charge $\frac{5}{3}e$, and the index ``$1$'' labels one of three families of quarks.
The other two left-handed quark families must come in an antitriplet representation of $SU(3)_L$ with the following content (again, right-handed quarks are singlets),
\begin{eqnarray}
&&Q_{i_L}^T=
\left (
%\begin{array}{c}
d^{\prime}_i\,, %\\
-u^{\prime}_i \,, %\\
J^{\prime}_i
%\end{array}
\right )^T_L \sim (\mbox{{\bf 3}},\mbox{${\bf 3^*}$},-\frac{1}{3}),\nonumber \\
u^{\prime}_{i_R}\sim(\mbox{{\bf 3}},&&\mbox{{\bf 1}},\frac{2}{3}),\,\,\,\,\,
d^{\prime}_{i_R}\sim(\mbox{{\bf 3}},\mbox{{\bf 1}},-\frac{1}{3}),\,\,\,\,\,J^{\prime}_{i_R}\sim(\mbox{{\bf 3}},\mbox{{\bf 1}},-\frac{4}{3}),
\label{quarks2-3} 
\end{eqnarray}
with $u^{\prime}_{i}$ and $d^{\prime}_{i}$ being the standard quarks and 
$J^{\prime}_{i}$  the exotic ones with electric charge $-\frac{4}{3}e$. Here the index $i=1,2$ represents the two remaining families. 

A curious outcome for this minimal 3-3-1 model is the presence of exotic quarks with electric charges $\frac{4}{3}e$ and $\frac{5}{3}e$. It is possible to show that these exotic quarks also carry two units of lepton number; in other words, they are fermionic bileptons, as we will discuss later. 
In addition to the standard gauge bosons, the model has five new ones, namely, a new $Z^{\prime}$, two simply charged gauge bosons $V^{\pm}$, and two doubly charged gauge bosons $U^{\pm \pm}$. We also call attention to the fact that these four charged gauge bosons carry two units of lepton number each, and are thus called bilepton gauge bosons. We stress that both, the exotic quarks and the doubly charged gauge bosons, constitute genuine and distinguishable signatures of the model.

Concerning the scalar sector of minimal the 3-3-1 model that provides spontaneous breaking of electroweak symmetry, we recall that in the original version of the model, three scalar triplets and one scalar sextet were necessary to generate the fermion masses. They are,
\begin{eqnarray}
\chi^T =
\left (
%\begin{array}{c}
\chi^- \,, %\\
\chi^{--} \,, %\\
\chi^0
%\end{array}
\right )^T,\,\,\,\,\,\rho^T &=&
\left (
%\begin{array}{c}
\rho^+ \,, %\\
\rho^0 \,, %\\
\rho^{++}
%\end{array}
\right )^T,\,\,\,\, 
\eta^T =
\left (
%\begin{array}{c}
\eta^0 \,, %\\
\eta^- \,, %\\
\eta^+
%\end{array}
\right )^T,\,\nonumber \\
S &=&
\left (
\begin{array}{lcr}
\sigma^0_1 & \frac{h_2^-}{\sqrt{2}} & \frac{h_1^+}{\sqrt{2}} \\
\frac{h_2^-}{\sqrt{2}} & H_1^{--} & \frac{\sigma_2^0}{\sqrt{2}} \\
\frac{h_1^+}{\sqrt{2}} & \frac{\sigma_2^0}{\sqrt{2}} & H^{++}_2
\end{array}
\right ).
\label{scalarsector} 
\end{eqnarray}
These scalars have the following transformation properties under the gauge group $SU(3)_C\otimes SU(3)_L\otimes U(1)_N$, $\chi \sim$ ({\bf 1},{\bf 3},-1),  $\rho \sim$ ({\bf 1},{\bf 3},1), $\eta \sim$ ({\bf 1},{\bf 3}, 0) and $S \sim$ ({\bf 1},{\bf 6},0). 
The neutral components of the triplets develop vacuum expectation values (VEVs), $\langle \eta^0\rangle = v_\eta$, $\langle \rho^0\rangle = v_\rho$, $\langle \chi^0\rangle = v_\chi$, so that we have the correct pattern of symmetry breaking, with $v_\chi$ being responsible for the $SU(3)_L\otimes U(1)_N$ breaking to $SU(2)_L\otimes U(1)_Y$ symmetry, while $v_\eta$ and $v_\rho$ combine to break $SU(2)_L\otimes U(1)_Y$ to the electric charge symmetry, $U(1)_{QED}$.
The necessity of a sextet scalar in the model is due to the fact that three triplets only are not sufficient to generate the correct masses of all fermions. The Yukawa Lagrangian in this case is,
\br
{\cal L}_Y &=& \frac{1}{2}G_{ab}{\bar f^C_{aL}} S^* f_{bL} +\l_1 {\bar Q_{1_L}}\chi J^\prime_{1_R} + \l_{ij} {\bar Q_{i_L}}\chi^* J^\prime_{j_R} +\l^\prime_{1a}{\bar Q_{1_L}}\rho d^\prime_{a_R}
\nonumber \\ 
&+& \l^\prime_{ia}{\bar Q_{i_L}}\rho^* u^\prime_{a_R} + \l^{\prime\prime}_{1a}{\bar Q_{1_L}}\eta u^\prime_{a_R} + \l^{\prime\prime}_{ia}{\bar Q_{i_L}}\eta^* d^\prime_{a_R} + h.c.
\label{yuka}
\er
In fact, from the above Yukawa Lagrangian, Eq.~(\ref{yuka}), we see that the three scalar triplets provide the correct masses of the quarks only. 
No coupling of leptons with the triplets is allowed~\footnote{Indeed, there is an unpleasant Yukawa coupling $G^\prime_{ab}\epsilon_{ijk}\bar f^C_{iaL} \eta_j f_{kbL}$ (here the indexes $i,\,j,\,k$ correspond to the components inside a triplet), which leads to unrealistic charged lepton masses~\cite{correctmasses}, that is always avoided through some  set of discrete symmetry. In our case it will be avoided by a $Z_3$ discrete symmetry to be imposed later.}, so to generate the lepton masses at least one sextet of scalars has to be added to the model~\cite{correctmasses}. 
Its presence does not guarantee a tree level Majorana mass term for the neutrinos though, unless the first component of the sextet acquires a nontrivial VEV. Otherwise, more ingredients have to be added or some radiative mechanism must be developed to appropriately generate neutrino masses. Observe that the lepton mass term in Eq.~(\ref{yuka}), seems to imply explicit lepton number violation, which is avoided at the Lagrangian level by assigning two units of lepton number to the $\s_1^0$ triplet component. The violation of lepton number can occur at the vacuum level when this field develops a VEV. It is this lepton number assignment that, through the lepton number conserving scalar potential~\cite{correctmasses}, implies nonzero lepton number to other scalars, which couple to the exotic quarks and to new gauge bosons, also doubly charged under lepton number, the so-called bileptons. This is a peculiar characteristic of some 3-3-1 models.	

Considering only one scalar sextet, $S$, the masses of all leptons, including neutrinos, arise from the common Yukawa coupling in Eq.~(\ref{yuka}),
\begin{eqnarray}
\frac{1}{2}G_{ab}{\bar f^C_{aL}} S^* f_{bL}\,.
\label{YukawaS}	
\end{eqnarray}
The sextet $S$ has two neutral components, $\sigma^0_1$ and $\sigma^0_2$, and when both develop nonzero VEV's, $v_{\sigma_1}$  and $v_{\sigma_2}$, we get nondiagonal neutrinos and charged lepton mass matrices given by the following expressions,
\be 
m_{\nu_{ab}}=G_{ab} v_{\sigma_1}\,\,\,\mbox{and}\,\,\, m_{l_{ab}}=G_{ab} v_{\sigma_2}\,. 
\label{mnue}
\ee
As we will explain later, this common Yukawa coupling for charged and neutral leptons is troublesome as far as the neutrino oscillation phenomena are concerned.

Among the most interesting predictions of the minimal 3-3-1 model is the one concerning the Weinberg mixing angle $\theta_W$, which arises from the kinetic term of the gauge bosons and is expressed by the following relation,
\begin{eqnarray}
\frac{g^{\prime}}{g}=\frac{S_W}{\sqrt{1-4S^2_W}}\,,
\label{g-gNrelation}	
\end{eqnarray}
where $S_W=\sin \theta_W$, $g$ is the $SU(3)_L$ coupling and $g^{\prime}$ is the $U(1)_N$ coupling. This relation shows us that a Landau pole exists for the theory and, in order to keep the theory inside the perturbative regime, a bound on this mixing angle is obtained, $S^2_W <0.25$. Translating this in terms of an energy scale, it was pointed out in Ref.~\cite{landaupole} that the perturbative regime of the model persists until about a few TeV. In other words, the highest energy scale where the model loses its reliable prediction power is about 4-5 TeV, and it can certainly be regarded as an effective model, meaning that above such a scale the underlying fundamental theory has to be called in. In view of this, the main problem that can be posed to the minimal 3-3-1 model is how is it possible to obtain light neutrino masses through a seesaw mechanism if the highest energy scale allowed by the theory is only around a few TeV? Finally, in the face of such a small scale for the underlying theory, the minimal 3-3-1 model is plagued by dangerous effective operators that could engender proton decay for example, the most significant operator given by,
\be
\frac{C_1\epsilon_{ijk}}{\Lambda^3}\overline{(Q_{1i_L})^c} f_{1j_L} \chi_k \overline{(u_{1_R})^c} d_{1_R} + {\mbox h.c.}\,,
\label{pdecay}
\ee
where $C_1$ is a dimensionless coupling and $\Lambda$ is the underlying fundamental theory scale, which is supposed to be around $5$~TeV. Here, $i,j,k=1,2,3$ are $SU(3)_L$ flavor indices and we are omitting the color indices which are implicit to be contracted with the antisymmetric structure constants forming a singlet under the color group.  We suppose that the first family is the one that contains the ordinary up and down quarks; for this reason, we have to take the $Q_{1_L}$ quark triplet as above in order to combine two up quarks and one down quark in a proton. 
One of the proton decay interactions produced by this dimension-7 operator is,
\be
\frac{C_1 v_\chi}{\Lambda^3} \bar{u^\prime}_{L}^c {e_L} \bar{u^\prime}_{R}^c d^\prime_{L} + {\mbox h.c.}\,,
\label{dim10}
\ee
which is responsible for proton decay through the transition of a $ud$
($uu$) pair to a $e{+}\bar{u}$ ($e{+}\bar{d}$) pair, $p\rightarrow e^+ + \pi^0 $. Making reasonable assumptions on the VEV's, $v_\eta\approx 100$~GeV, $v_\chi\approx 1$~TeV, and taking $\Lambda\approx 5$~TeV, we obtain that the ratio $\frac{C_1 v_\chi}{\Lambda^3}$ is about $10^{-8}$~GeV$^{-2}$ for $C_1$ of the order of unit. This result would imply a proton lifetime of only some $10^{-8}$~s. This astoundingly contradicts the experiment, and to avoid such a result, the dimensionless coupling constant $C_1$ should be unnaturally tiny in the minimal 3-3-1 model. We show in the next section how to circumvent the problem of neutrino mass as well as the fast proton decay with a smaller scalar content in the minimal 3-3-1 model with additional discrete symmetries.

\section{ The truly minimal 3-3-1 model}
\label{sec2}
In this section we propose a modification of the minimal 3-3-1 model which  mainly  solves the problem of getting light neutrinos at the eV scale while keeping the observed feature of neutrino oscillation. Besides, we also show a way out for the above-mentioned proton decay puzzle in the minimal 3-3-1 model. In this sense we are going to obtain as a byproduct what we may call a truly minimal version of the model.

\subsection{Lepton masses}

We start this section by showing that a unique scalar sextet is not enough to generate the correct masses of all leptons of the model. The reason is simple, notice that the texture of $m_\nu$  is the same as $m_l$ because both are proportional to the same matrix $G_{ab}$. Automatically, the rotating mixing matrix $U$ that diagonalizes $m_\nu$ will diagonalize $m_l$ too. Thus, the lepton mass eigenstates $\hat{l}$ are related to the symmetry eigenstates $l$ through the following rotation ,
\begin{eqnarray}
\hat{\nu}_{l_L} =U_{la} \nu_{a_L}\,\,\,\,\,\,\mbox{and}\,\,\,\,\,\, \hat{e}_{l_L}=U_{la}e_{a_L}\,,
\label{eigenstates}
\end{eqnarray}
where $a=1,2,3$. Consequently, the charged current will always be diagonal,
\begin{eqnarray}
\frac{g}{\sqrt{2}}\bar{\hat{\nu}}_L \gamma^\mu \hat{e}_L W^+_\mu \rightarrow \frac{g}{\sqrt{2}}\bar \nu_L \gamma^\mu e_L W^+_\mu\,,
	\label{chargedcurrent}
\end{eqnarray}
which goes against the recent atmospheric and solar neutrino oscillation experiments. As an immediate consequence, three scalar triplets and one scalar sextet are not sufficient to explain the masses of all fermions of the model. There are, in the literature, many suggestions for solving  this problem  but all of them require the enlargement of the particle content of the model~\cite{neutrinos331}.
To complicate matters even more, recall that the highest energy scale available to the perturbative theory is about 5~TeV. Thus, even if we enlarge the scalar sector of the model with the intent of obtaining the correct mass terms for the neutrinos through some kind of seesaw mechanism, we still face the problem of how to appropriately generate small neutrino masses in the eV range. Such a small underlying energy scale just does not comply.  

In order to fully understand this peculiarity in the problem of the smallness of neutrino masses in the minimal 3-3-1 model, let us  forget for a while the scalar sextet and generate the neutrino mass through an effective dimension-5 operator, 
\begin{eqnarray}
\frac{h}{\Lambda}(\bar f^C_L \eta^*)(\eta^{\dagger}f_L),
\label{effectiveD-5}
\end{eqnarray}
where $h$ is a dimensionless coupling and $\Lambda$ is the mass scale of the underlying fundamental theory. This operator yields the following mass formula for the neutrinos, 
\begin{eqnarray}
m_\nu= \frac{h v^2_\eta}{\Lambda}\,.
\label{D-5masses}
\end{eqnarray}
Notice that, for $v_\eta \approx 10^2$~GeV and $\Lambda=5$~TeV, we get $m_\nu = 10\,h$~GeV. In other words,  even if we implement some appropriate mechanism to generate tiny neutrino masses, the non-negotiable presence of an effective dimension-5 operator is predicting heavy neutrinos, washing out the lightness of the neutrino mass. Of course, the addition of more scalars might be enough to generate different mass textures and lead to the observed mixing in the charged current, but clearly it would be helpless in dealing with the effects of such a dimension-5 operator. So, concerning neutrinos,  this is the present undesirable status of the minimal 3-3-1 model. 

Let us now provide a solution to this problem. Our basic idea is to eliminate the scalar sextet and try to generate the masses of the leptons through effective operators.  These operators should emerge from the underlying fundamental theory anyway, but we will not be concerned about the details of such a theory. What matters to us in this work is to investigate its low energy effects through effective operators. In avoiding the scalar sextet, the particle content of the model involves the leptons and quarks in Eqs.~(\ref{leptoniccontent}), (\ref{quarks1}) and (\ref{quarks2-3}), and the three triplets $\eta$, $\rho$  and $\chi$  in Eq.~(\ref{scalarsector}). 

In order to get rid of the effective dimension-5 operator in Eq.~(\ref{effectiveD-5}) that generates heavy neutrino masses, we evoke the discrete symmetry $Z_3$ with the following representation for the scalar and fermion fields,
\begin{eqnarray}
&&	\eta \rightarrow  e^{4i\pi/3}\eta ,\,\rho \rightarrow e^{-4i\pi/3}\rho , \,\chi \rightarrow e^{2i\pi/3} \chi , \nonumber \\
&&f_{l_L} \rightarrow e^{2i\pi/3}f_{l_L} ,\,d_{a_R} \rightarrow e^{4i\pi/3} d_{a_R} ,\,u_{a_R} \rightarrow e^{-4i\pi/3} u_{a_R},\nonumber \\
&&J_{1_R} \rightarrow e^{-2i\pi/3} J_{1_R} ,\,J_{i_R} \rightarrow e^{2i\pi/3} J_{i_R},
\label{Z3symmetry}
\end{eqnarray}
and all the other fields transform trivially under the $Z_3$ symmetry.  It is important to point out that all the original Yukawa couplings that lead to the correct mass terms for all quarks, Eq.~(\ref{yuka}), are maintained in the presence of this $Z_3$ symmetry, so we do not worry about this sector. Let us focus then on the lepton masses, starting with the neutrinos.

It is amazing that such $Z_3$ symmetry allows, as the first dominant effective operator that generates the neutrino mass, exactly a dimension-11 operator,
\begin{eqnarray}
\frac{h^\nu }{\Lambda^7 }(\bar f^C_L \eta*)( \rho \chi \eta)^2(\eta{\dagger}f_L ),
\label{D11neutrinos}
\end{eqnarray}
where $h^\nu$ is a dimensionless coupling. In this equation the term $(\rho \chi \eta)$ is the antisymmetric singlet combination under $SU(3)_L$.
It is easy to see that lower dimension terms are forbidden since we need to couple the lepton triplet to the $\eta$ triplet in order to select the neutrino when the $\eta^0$ acquires a VEV,
\be
(\bar f^C_L \eta*)(\eta{\dagger}f_L )\,
\label{lowest}
\ee
which is gauge invariant.
However, the $Z_3$ symmetry demands that we multiply this resulting term by some combination of scalars that transforms as $e^{4i\pi/3}$ under $Z_3$. In this way, no power of hermitian terms of the kind $\phi^\dagger \phi$ (where $\phi$ is any of the scalar triplets) will do the job alone. The only combination of triplets that can be used to get the lowest dimension operator is the gauge invariant antisymmetric combination $\epsilon_{ijk} \rho_i \chi_j \eta_k$ and its powers. The second power in this term is the least we can add that possesses the right $Z_3$ quantum number.

When $\eta$, $\rho$  and $\chi$ develop their respective VEVs, the dimension-11 operator in Eq.~(\ref{D11neutrinos}) generates the following  expression for the neutrino masses,
\be
m_\nu=\frac{h^\nu}{\Lambda^7} v_\eta^4 v_\rho^2 v_\chi^2\,.
\label{neutrinomassexpression}
\ee
That is a striking result if one observes that even for $\Lambda$ around $v_\chi$, the operator above is suppressed enough to generate small masses for the neutrino. For instance, in being conservative and taking, as before, $v_\chi \approx 1$~TeV,
$v_\eta \approx 10^2$~GeV, $\Lambda = 5$~TeV and $v_\rho \approx 10$~GeV~\footnote{We remember that $v_\eta$ and $v_\rho$ combine to give the Standard Model VEV, such that this choice for the value of $v_\rho$ is quite reasonable.},
we obtain a prediction for the neutrino masses,	$m_\nu \approx 0.1\, h^\nu$~eV, meaning that
a $\Lambda$ around a few TeV is perfectly compatible with neutrino masses at the sub-eV scale, which represents an astonishing achievement for the model. 

Now let us  focus on the charged leptons. In the presence of the $Z_3$ symmetry the dominant effective operator that generates their masses is,
\begin{eqnarray}
	\frac{h^l}{\Lambda}(\bar f^C_L \rho^*)(\chi^{\dagger} f_L) + \frac{h^{\prime l}}{\Lambda}(\bar f^C_L \chi^*)(\rho^{\dagger} f_L) +H.c.\,
	\label{D5chagedleptons}
\end{eqnarray}
Here the parameters $h^l$ and $h^{\prime l}$ are dimensionless couplings. After spontaneous symmetry breaking, this term generates the following expression for the charged lepton masses,
\begin{eqnarray}
m_l=\frac{h^l+h^{\prime l}}{\Lambda}v_\rho v_\chi\,.
\label{leptomasses}
\end{eqnarray}
Again, this is formidable because since $\Lambda$ is around $5\,v_\chi$, we have $m_l \approx (h^l + h^{\prime l}) v_\rho/5$. This is very close to the SM expression for the charged lepton masses, with less fine-tuning in the couplings. To check this, by taking the same set of values for the VEVs and $\Lambda$ assumed above, we obtain the following prediction for the charged lepton masses, $m_l\approx 2(h^l + h^{\prime l})\mbox{GeV}$.
There is an evident gain in relation to the SM prediction, where $m_l \approx 10^2\, y^l$~GeV, and $y^l$ is the SM Yukawa couplings for the charged leptons. In our case, considering charged leptons in a diagonal basis, we just need couplings in the range $10^{-4}-10^{-1}$ to get the correct masses for all charged leptons. This has to be contrasted with the SM scenario that requires charged lepton Yukawa couplings from $10^{-6}-10^{-2}$, and the situation is worsened for the neutrinos if one includes right-handed singlet neutrinos in the SM to give them a Dirac mass. Thus, our model also represents a small improvement concerning the large hierarchy of Yukawa couplings for leptons in the SM.
Besides, the problem of a diagonal charged current in the minimal model can be solved if we take into account that the effective couplings in Eq.~(\ref{neutrinomassexpression}) are nondiagonal, since they are unrelated to the charged lepton couplings in Eq.~(\ref{leptomasses}). The neutrino oscillation pattern may then be recovered by a judicious choice of these couplings.

Considering all this, i.e., that our model needs three scalar triplets only and is still able to generate the correct masses for all fermions, including neutrinos at the eV range,  without jeopardizing the mixing in the electroweak charged current, we claim that this version of the model developed here with a $Z_3$ symmetry is, in fact, the truly minimal 3-3-1 model. Next we approach the problem of fast proton decay in the truly minimal 3-3-1 model.

\subsection{Proton decay}
Although we have shown how to make the minimal 3-3-1 model consistent with the observed leptonic spectrum, our  proposal suffers from the same disease as the minimal 3-3-1 model, namely, fast proton decay when no unnaturally tiny coupling constants are allowed. In the truly minimal 3-3-1 model the dangerous operator of lowest dimension is the dimension-8 operator, 
\be
\frac{C_2\epsilon_{ijk}\epsilon_{lmn}}{\Lambda^4}(\overline{Q_{1i_L}^c} f_{1j_L} \chi_k)  (\overline{Q_{1l_L}^c} Q_{1m_L} \chi_n) + {\mbox h.c.}\,,
\label{pdecay2}
\ee
where, $i,j,k,l,m,n=1,2,3$ are $SU(3)_L$ indices and again we assumed an antisymmetric singlet combination in color space. $C_2$ is a dimensionless coupling, with $\Lambda$ representing the underlying fundamental energy scale. Observe that we have four fermion fields transforming as triplets under $SU(3)_L$, which makes the antisymmetric combinations taken above, the most economical choice, invariant under the gauge group and the $Z_3$.
One of the proton decay interaction terms that this dimension-8 operator produces is (omitting the antisymmetric tensors),
\be
\frac{C_2 v_\chi^2}{\Lambda^4} \overline{u_{L}^{\prime c}} {e_L} \overline{u_{L}^{\prime c}} d^\prime_{L} + {\mbox h.c.}\,,
\label{dim13}
\ee
which, using the same assignments for the parameters as before, yields an effective coupling for $C_2 \sim 1$ that is of order of $4\times 10^{-5}$~GeV$^{-2}$, and a proton lifetime around $10^{-7}$~s, still far too short. The simplest solution to this problem is to admit an additional discrete symmetry, $Z_2$, working on the following fields,
\be
Q_{a_L} \rightarrow - Q_{a_L}\,,\,\,\,\,Q_{a_R}\rightarrow -Q_{a_R}\,,
\label{z2}
\ee
with $Q_{a_R}$ representing all right-handed singlet quarks, ordinary and exotic ones, with $a=1,2,3$. All the remaining fields are even under this $Z_2$. It can be easily seen that the dimension-8 operator in Eq.~(\ref{pdecay2}) is forbidden together with all relevant proton decay operators which were allowed before this symmetry since they involve an odd number of quark fields. Clearly, this eliminates the problem from the beginning for any model, but here it is a necessity (as well as for the minimal 3-3-1 model), differently from the SM, for example.

\section{Summary}
\label{sec3}
In conclusion, we proposed a truly minimal 3-3-1 model for the electroweak interactions where the scalar content is composed  of three scalar triplets only. The essence of the idea is to impose a $Z_3$ symmetry with a particular representation that forbids Yukawa couplings at tree level involving the lepton triplets. As a consequence, charged lepton masses arise from effective dimension-5 operators while neutrino masses arise from effective dimension-11 operators. Remarkably, for typical values of VEVs of the theory, we obtain neutrino and charged lepton masses in the right energy range, namely neutrino masses at the sub-eV scale  and charged lepton masses at the GeV scale, with naturally occurring Yukawa couplings, meaning no (or little) fine-tuning when compared to the SM. Besides, we have shown that the minimal 3-3-1 model faces a serious problem concerning proton decay, caused by unavoidable dimension-7 effective operators. This undesirable feature is also present in our model through dimension-8 operators, which slightly alleviates the problem but still implies a fast proton decay. 
We choose to eliminate such operators at any order by imposing a $Z_2$ symmetry that kills any operator involving an odd number of quark fields, without jeopardizing any other term in the model Lagrangian that reproduces the well-known quark phenomenology.
%%%%%%%%%%%%%%%%%%%%%%%%%%%%%%%%%%%%%%%%%%%%%%%%%%%%%%%%%%%%%%%%%%%%%

\textit{Acknowledgments}
This work was supported by Conselho Nacional de Pesquisa e
Desenvolvimento Cient\'{i}fico- CNPq and Coordena\c c\~ao de Aperfei\c coamento de Pessoal de N\'{i}vel Superior - CAPES.

%%%%%%%%%%%%%%%%%%%%%%%%%%%%%%%%%%%%%%%%%%%%%%%%%%%%%%%%%%%%%%%%%%%%%%%%%%%%%%%%%%%%%%%%%%%

%%%%%%%%%%%%%%%%%%%%%%%%%%%%%%%%%%%%%%%%%%%%%%%%%%%%%%%%%%%%%%%%


\begin{thebibliography}{99}
%%%%%%%%%%%%%%%%%%%%%%%%%%%%%%%%%%%%%%
\bibitem{osc}
Super-Kamiokande Collaboration (Y.~Fukuda {\it et al.}), Phys. Rev. Lett. 
{\bf 81}, 1562 (1998)
%
\bibitem{DM}
Supernova Cosmology Project, (S. Perlmutter {\it et al.}), Astrophys. J.{\bf 517} (1999), 565. 
%
\bibitem{pal}
P. B. Pal, Phys. Rev. D{\bf 52} (1995) 1659;
A. G. Dias, V. Pleitez and M. D. Tonasse, Phys. Rev. D{\bf 67} (2003) 095008; 
A. G. Dias, C. A. de S. Pires, P. S. Rodrigues da Silva, Phys. Rev. D{\bf 68}(2003) 115009.
A. G. Dias, V. Pleitez and M. D. Tonasse, Phys. Rev. D{\bf 69}, (2004) 015007. 
%
\bibitem{ECQ}
C. A. de S. Pires and O. P. Ravinez, Phys. Rev. D {\bf 58}(1998), 035008;
C. A. de S. Pires, Phys. Rev. D {\bf 60}(1999), 075013.
%
\bibitem{DM331}
D. Fregolente, M. D. Tonasse, Phys. Lett. B{\bf 555} (2003), 7-12;
H. N. Long, N. Q. Lan, Europhys. Lett. {\bf 64} (2003), 571;
S. Filippi, W. A. Ponce, L. A. Sanchez, Europhys. Lett. {\bf 73} (2006), 142; 
C. A. de S. Pires, P. S. Rodrigues da Silva, JCAP {\bf 0712} (2007), 012. 
%

\bibitem{earlymodels}
For a list of references although incomplete, see: J. Schechter and Y. Ueda, Phys. Rev.  D {\bf 8}, 484 (1973); J.  G. Segr\`e and J. Weyers, Phys. Lett.  B {\bf 65}, 243 (1976); H. Fritzch and P. Minkowski, Phys. Lett. 
B {\bf 63}, 204 (1976); B. W. Lee and S. Weinberg, Phys. Rev.  Lett. {\bf 38}, 1237 (1977); 
B. W. Lee and R. E. Shrock, Phys. Rev. D {\bf 17}, 2410 (1978);  P. 
Langacker, G. Segr\`e and M. Golshani, Phys. Rev.  D {\bf 17}, 1402 (1978); M. Singer, J. W. 
F. Valle and J. Schechter, Phys. Rev.  D {\bf 22}, 738 (1980).
%
\bibitem{ppf}
F. Pisano and V. Pleitez, Phys. Rev.  D {\bf 46 }(1992), 410; 
 P. H. 
Frampton, Phys. Rev. Lett. {\bf 69} (1992), 2889.
%
\bibitem{footpp}
R. Foot, H. N. Long and T. A. Tran, Phys. Rev. D{\bf 50},
(1994) R34; J. C. Montero, F. Pisano and V. Pleitez, Phys. Rev. D{\bf 47} (1993), 2918. 
 
%
\bibitem{landaupole}
A. G. Dias, R. Martinez, V. Pleitez,  Eur. Phys. J. C{\bf 39}(2005), 101. see also,  A. G. Dias, V. Pleitez, Phys. Rev. D{\bf 80}(2009), 056007.
%
\bibitem{baryonv} S.~Weinberg, Phys. Rev. Lett. {\bf 43} (1979), 1566.
%
\bibitem{correctmasses}
R. Foot, O. F. Hernandez, F. Pisano, V. Pleitez, Phys. Rev. D{\bf 47}(1993), 4158. 
%
\bibitem{neutrinos331}
Y. Okamoto, M. Yasue, Phys. Lett. B{\bf 466} (1999), 267;
T. Kitabayashi, M. Yasue, Nucl. Phys. B{\bf 609} (2001), 61;
J. C. Montero, C. A. de S. Pires, V. Pleitez, Phys. Rev. D{\bf 66} (2002), 113003. 


%%%%%%%%%%%%%%%%%%%%%%%%%%%%%%%%%%%%%%%%%%%%%%%%
\end{thebibliography}
\end{document}